\documentclass[twocolumn,showpacs,preprintnumbers,amsmath,amssymb,floatfix,nofootinbib]{revtex4-2}

\usepackage{graphicx}
\usepackage{dcolumn}
\usepackage{bm}
\usepackage{color}
\usepackage{ulem}

\usepackage{epstopdf}

\begin{document}

\title{Coherent spin dynamics of electrons and holes photogenerated with large kinetic energy in lead halide perovskite crystals}

\author{Evgeny~A.~Zhukov$^{1}$, Dmitri~R.~Yakovlev$^{1}$, Erik~Kirstein$^{1}$, Nataliia~E.~Kopteva$^{1}$, Oleh Hordiichuk$^{2,3}$,  Maksym V. Kovalenko$^{2,3}$, and Manfred~Bayer$^{1,4}$}

\affiliation{$^1$Experimentelle Physik 2, Technische Universit\"at Dortmund, 44227 Dortmund, Germany}
\affiliation{$^2$Laboratory of Inorganic Chemistry, Department of Chemistry and Applied Biosciences, ETH Z{\"u}rich, CH-8093 Z{\"u}rich, Switzerland}
\affiliation{$^3$Laboratory for Thin Films and Photovoltaics, Department of Advanced Materials and Surfaces, Empa - Swiss Federal Laboratories for Materials Science and Technology, CH-8600 D{\"u}bendorf, Switzerland}
\affiliation{$^4$Research Center FEMS, Technische Universit\"at Dortmund, 44227 Dortmund, Germany}

\date{\today}

\begin{abstract}
The coherent spin dynamics of electrons and holes are studied in a FA$_{0.9}$Cs$_{0.1}$PbI$_{2.8}$Br$_{0.2}$ perovskite bulk crystal, using time-resolved  Kerr ellipticity in a two-color pump-probe scheme. The probe photon energy is tuned to the exciton resonance, while the pump photon energy is detuned from it up to 0.75~eV to higher energies. The spin-oriented electrons and holes photogenerated with significant excess kinetic energy relax into states in vicinity of the band gap, where they undergo Larmor precession in an external magnetic field. At cryogenic temperatures down to 1.6~K, the spin dephasing time reaches the nanosecond range. During energy relaxation, the carrier spin relaxation is inefficient and only happens when the carriers become localized. In experiments with two pump pulses, all-optical control of the amplitudes and phases of the electron and hole spin signals is achieved in the additive regime by varying the intensities of the pump pulses and the time delay between them.

\end{abstract}


\maketitle


\section{Introduction}
\label{sec:1}

The lead halide perovskite semiconductors form an emerging class of materials with interesting optical and electronic properties, making them favorable for photovoltaics and optoelectronics applications~\cite{Vardeny2022_book,Vinattieri2021_book,jena2019}. Their band structure considerably differs from that of conventional III-V and II-VI semiconductors, suggesting that the perovskites are interesting model materials for spin physics relevant for spintronics and spin-orbitronics applications~\cite{Vardeny2022_book,Privitera2021,wang2019,kim2021}. In turn, the spin physics sheds light on details of the band structure and the underlying crystal symmetry, as the Land\'e $g$-factors of the charge carriers, as well as their spin polarization and spin dynamics are highly dependent on them~\cite{kirstein2022nc,XOO2024,XOO2024_all}.

Time-resolved Faraday/Kerr rotation (TRFR or TRKR) is a powerful tool to study the spin properties of semiconductors and their nanostructures~\cite{Awschalom2002,Yakovlev_ch6_2017}. The technique exploits optical spin orientation of charge carriers and/or excitons by circularly polarized pump pulses and detection of their spin dynamics via Faraday or Kerr rotation of linearly polarized probe pulses that can be delayed in time. In an external magnetic field, Larmor spin precession can be detected which provides information on the involved carrier $g$-factors and their spin dephasing times $T_2^*$. The technique has been implemented 30 years ago~\cite{Baumberg1994,Zheludev1994} and since that time has been extended by various modifications, for example, leading to a two-color (pump and probe have different photon energies)~\cite{Zhukov2007} or two-pump pulse approach (pump pulses are delayed in time in respect to each other and also may differ in photon energies and in polarization)~\cite{Akimoto1998,Greilich2009,Phelps2009,Zhukov2010}, the extended pump-probe method to measure long longitudinal spin relaxation time $T_1$~\cite{Belykh2016}, resonant spin amplification (RSA) and spin mode-locking~\cite{Kikkawa1998,Greilich2006,Yugova2012,kirstein2023_SML}, polarization recovery in a longitudinal magnetic field~\cite{Smirnov2020}, etc. All these schemes greatly contribute to gaining a detailed understanding of the spin properties and especially the spin dynamics of charge carriers in semiconductors.

Time-resolved Faraday/Kerr rotation has been successfully used to study the coherent spin dynamics of electrons and holes in the lead halide perovkite semiconductors. Among the studied samples are bulk crystals of CsPbBr$_3$~\cite{belykh2019,Huynh2022cspbbr3}, FA$_{0.9}$Cs$_{0.1}$PbI$_{2.8}$Br$_{0.2}$~\cite{kirstein2022am}, MAPbI$_3$~\cite{Kirstein_MAPI_2022,Huynh2022mapi}, FAPbBr$_3$~\cite{Kirstein_2024_FAPbBr3}, MAPbBr$_3$~\cite{Huynh2024},  and polycrystalline films of CsPbBr$_3$~\cite{Grigoryev2021,Jacoby2022}, MAPb(Cl,I)$_3$~\cite{odenthal2017}, MAPbI$_3$~\cite{garcia-arellano2021,garcia-arellano2022}, and FAPbI$_3$~\cite{Lague2024}. It has been established that the electron and hole Land\'e $g$-factors follow a universal dependence on the band gap energy for the whole class of lead halide perovskites~\cite{kirstein2022nc}. Also the spin relaxation times at cryogenic temperatures differ not drastically for various materials lying in the range of about $0.5-11$~ns for the spin dephasing time $T_2^*$ and $30-250$~ns for the longitudinal spin relaxation time $T_1$. An overview of the characteristic spin relaxations times measured for these materials can be found in ref.~\onlinecite{Kirstein_2024_FAPbBr3}.

The two-pump and two-color approaches were applied to CsPbBr$_3$ nanocrystals~\cite{Lin2023} and CsPb(Cl,Br)$_3$ nanocrystals~\cite{kirstein2023_SML}, but so far not for bulk perovskites. In this paper we present two-color measurements motivated by our recent results on optical orientation of excitons and charge carriers in perovskite crystals. A large optical orientation degree of 85\% was reported for FA$_{0.9}$Cs$_{0.1}$PbI$_{2.8}$Br$_{0.2}$ crystals, which is stable against energy detuning between the excitation photon energy and exciton resonance energy~\cite{XOO2024,COO2024}. The degree was independent of detuning up to 0.3~eV and then smoothly decreases to zero for the detunings reaching 1.0~eV. The reason for this finding is the absence of the Dyakonov-Perel spin relaxation mechanism in perovskite crystals with spatially inversion symmetry~\cite{Kepenekian2015,Kepenekian2017,XOO2024_all}. This suggests that in TRFR/TRKR experiments the efficient generation of spin coherence for localized carriers can be provided at large detunings. This allows one to obtain information on the carrier spin relaxation during energy relaxation. We examine that in the present paper using as test bed FA$_{0.9}$Cs$_{0.1}$PbI$_{2.8}$Br$_{0.2}$ crystals with in-depth studied spin properties~\cite{kirstein2022am,XOO2024,COO2024}.  We also use the two-pump pulses technique to test the ability for all-optical control of the electron and hole spin coherence in perovskite crystals.

In detail, we study the coherent spin dynamics of electrons and holes in FA$_{0.9}$Cs$_{0.1}$PbI$_{2.8}$Br$_{0.2}$ bulk crystal using the time-resolved Kerr ellipticity technique. We use both one-color and two-color schemes and implement one-pump or two-pump excitation for the spin control experiments.

\section{Results and Discussion}

\begin{figure*}[hbt]
\begin{center}
\includegraphics[width=16cm]{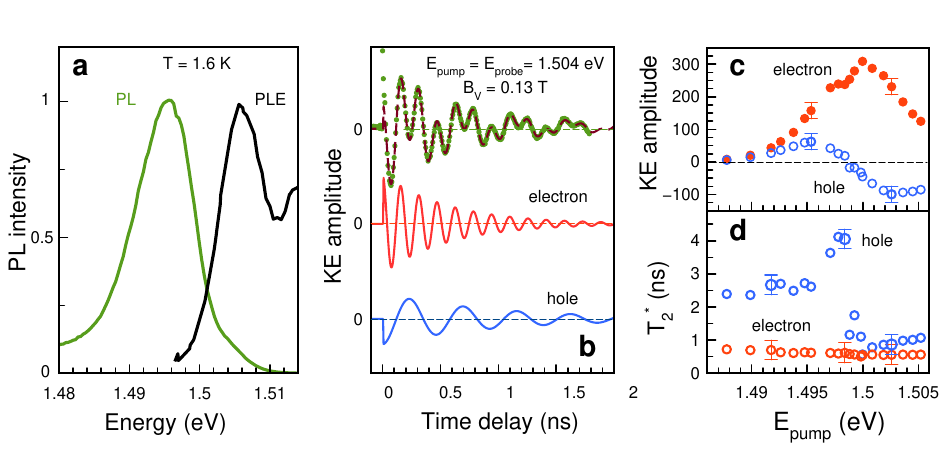}
\caption{Optical properties and spin dynamics measured for FA$_{0.9}$Cs$_{0.1}$PbI$_{2.8}$Br$_{0.2}$ crystal at $T=1.6$~K. (a) Photoluminescence spectrum (green line) measured using continuous-wave (cw) excitation at 1.653 eV photon energy.  Photoluminescence excitation spectrum (black line) detected at 1.495~eV under cw excitation.  (b) Kerr ellipticity dynamics (green circles) at $B_{\rm V}=0.13$~T measured at $E_{\rm pump} = E_{\rm probe}= 1.504$~eV. $P_{\rm pump}=7$~W/cm$^{2}$ and $P_{\rm probe}=1$~W/cm$^{2}$. Dashed brown line is a fit with eq~\eqref{n1} and parameters given in the text. The individual components for the electron (red line) and hole (blue line) spin dynamics obtained from the fit are shown below. Dashed lines indicate the corresponding zero levels.  Spectral dependences of the KE amplitude (c) and the spin dephasing time (d) of electrons (red circles) and holes (blue circles).}
\label{Fig:1}
\end{center}
\end{figure*}

The photoluminescence (PL) spectrum of the FA$_{0.9}$Cs$_{0.1}$PbI$_{2.8}$Br$_{0.2}$ crystal, measured at $T=1.6$~K temperature using continuous-wave excitation, is shown in Figure~\ref{Fig:1}a. The spectrum peaks at 1.495 eV and shows a weak shoulder at 1.506~eV corresponding to the exciton resonance, which can be well resolved in the PL excitation (PLE) spectrum (black line). The Stokes shifted PL is contributed by recombination of localized electrons and holes as well as bound excitons, for details see refs.~\onlinecite{kirstein2022am,COO2024}.

The population dynamics of the excitons and carriers are measured by time-resolved differential reflection, see Figure S1 in the Supporting Information. The dynamics have two characteristic times in the measured temporal range up to 5~ns. We assign the first time of about 0.1~ns to the recombination of excitons and the second of about 7~ns to the recombination of localized electrons and holes. The presence of localized electrons and holes, which are spatially separated from each other, is typical for lead halide perovskite crystals at cryogenic temperatures. In previous studies, it was evidenced by the multi-exponential recombination dynamics, with times ranging from nanoseconds to hundred microseconds, as well as by the coexistence of electron and hole spin beats in the spin dynamics measured in magnetic field~\cite{belykh2019,kirstein2022am}.

To study the coherent spin dynamics of electrons and holes we use a time-resolved pump-probe technique detecting the Kerr ellipticity (KE)~\cite{Yakovlev_ch6_2017,Glazov2010}. The used laser has a pulse duration of 1.5~ps with a pulse repetition period of 13.2~ns. Both pump-probe color schemes, the one-color scheme (pump and probe have the same photon energy) and the two-color scheme (the pump and probe energies are different), are applied. Details of these methods are given in the Supplementary Information S2. Most of the experiments presented below are performed at the temperature of $T=1.6$~K.

\subsection{Electron and hole spin coherence measured in one-color scheme}
\label{one-color}

We begin with the conventional one-color pump probe scheme, where the pump and probe energies are degenerate. A typical KE dynamics, measured at the photon energy of $E_{\rm pump} =E_{\rm probe}=1.504$~eV in the magnetic field of $B_{\rm V} = 0.13$~T applied in the Voigt geometry, is shown in Figure~\ref{Fig:1}b. The experimental dynamics, shown by the green symbols, is contributed by two Larmor precession frequencies, which are shown individually at the bottom of this figure. To decompose the KE signal into its components we use the following equation:
\begin{equation}
A_{\text{KE}}(t)=\sum_{i}{A_{0,i}\exp\left(-\frac{t}{T_{2,i}^*}\right)\cos(\omega_{{\rm L},i}t )}.
\label{n1}
\end{equation}
Here $A_{0,i}$ is the amplitude of the signal corresponding to the electron and hole components ($i=e,h$). The carrier  $g$-factors $g_{i}$ are directly linked to the Larmor precession frequencies and can be evaluated according to:
\begin{equation}
|g_{i}|=\frac{\hbar \omega_{{\rm L},i}}{\mu_{\rm B}B_{\rm V}}.
\label{n2}
\end{equation}
Here $\hbar$ is the Planck constant and $\mu_{\rm B}$ is the Bohr magneton.
The fit of the KE dynamics from Figure~\ref{Fig:1}b gives the following parameters for the electrons $T_{2,e}^*=0.6$~ns, $g_e=3.54\pm0.11$ and for the holes $T_{2,h}^*=1.1$~ns, $g_h=-1.19\pm0.05$ with $A_{0,e}/A_{0,h}=1.5$. The justification of the assignment of the faster precession to the electrons and of the slower one to the holes, as well as the clarification of the $g$-factor signs for the FA$_{0.9}$Cs$_{0.1}$PbI$_{2.8}$Br$_{0.2}$ crystal were discussed in refs.~\onlinecite{kirstein2022nc,kirstein2022am}.

By tuning the laser photon energy in the exciton spectral range of $1.480-1.505$~eV we find a clear resonance behavior. The associated KE dynamics and the spectral dependencies of spin parameters, see eq~\eqref{n1}, are collected in Figure~\ref{Fig:1S}. The Larmor precession frequencies, i.e. also the electron and hole $g$-factors, do not change in this spectral range (Figure~\ref{Fig:1S}c). The electron KE amplitude maximum is at 1.500~eV, which is slightly shifted to lower energy relative to the exciton resonance in the PL excitation spectra, see Figures~\ref{Fig:1}a,c. The hole KE amplitude spectral dependence has a line width similar to that for the electron, but with a dispersive shape, crossing zero at about 1.499~eV (Figure~\ref{Fig:1}c). Such dispersively-shaped dependence is not expected for KE signals~\cite{Glazov2010}, while it was recently reported also for holes and electrons in MAPbI$_3$ crystals~\cite{Kirstein_MAPI_2022}. The clarification of the responsible mechanism requires further investigation. The spin dephasing time of electrons is  $T_{2,e}^*\approx0.6$~ns being spectrally independent (Figure~\ref{Fig:1}d). The hole spin dephasing time changes from about $T_{2,h}^*\approx1$~ns at energies exceeding 1.500~eV rather abruptly to significantly larger values of about 2.5~ns and more at lower energies. This shows that the resident holes have different localization energies. 

For completeness we perform detailed study of the KE dynamics in the one-color regime at $E_{\rm pump} =E_{\rm probe}=1.503$~eV in magnetic fields up to 2.1~T and for temperatures in the range of $1.6-25$~K, see Supplementary Information Figure~\ref{Fig:3S}. Notable results are the strong decrease of the KE amplitude and shortening of the spin dephasing time with increasing temperature to 25~K.  The results are in line with what we reported for FA$_{0.9}$Cs$_{0.1}$PbI$_{2.8}$Br$_{0.2}$ in ref.~\onlinecite{kirstein2022am}. Together with the long spin dephasing times at $T = 1.6$~K, which considerably exceed the exciton recombination time, this clearly shows that the resident electrons and holes with long lifetimes provide the measured coherent spin dynamics.

\subsection{Spin coherence measured in two-color scheme}
\label{two-color}

In the two-color scheme the photon energies of the pump and probe are different. The probe is set to the exciton resonance, but the pump energy is detuned from it to larger values by $\Delta=E_{\rm pump}-E_{\rm probe}$. Spin-oriented electrons and holes are photogenerated with excess kinetic energy and (partial) spin relaxation may take place during the carrier energy relaxation toward the band gap. Measurements of the KE dynamics for different detunings give access to spin relaxation and the responsible mechanisms.

\begin{figure*}[hbt!]
\begin{center}
\includegraphics[width=14cm]{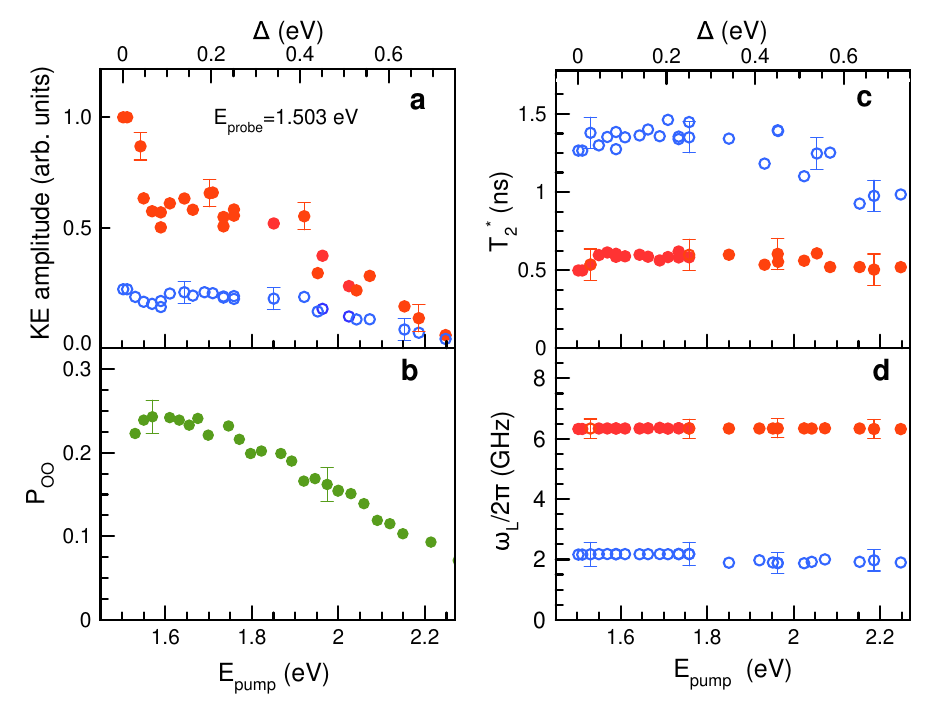}
\caption{Coherent spin dynamics in the FA$_{0.9}$Cs$_{0.1}$PbI$_{2.8}$Br$_{0.2}$ crystal measured in the two-color scheme with $E_{\rm probe} = 1.503$~eV. The pump energy detuning from the probe energy ($\Delta=E_{\rm pump}-E_{\rm probe}$) is given by the upper axis. Red symbols are for the electrons and blue symbols for the holes. $B_{\rm V} = 0.13$~T and $T=1.6$~K. The following dependences on the pump detuning are shown: (a) Kerr ellipticity amplitude, (c) spin dephasing time,  and (d) Larmor precession frequency. (b) Dependence of the optical orientation degree on the excitation energy, evaluated from time-integrated photoluminescence, detected at 1.504~eV. These data are taken from ref.~\onlinecite{COO2024}.}
\label{Fig:2}
\end{center}
\end{figure*}

Importantly, we find experimentally that in the two-color scheme the properties of the generated carrier spin coherence does not change considerably, even for large detunings. One can see that from the comparison of the sets of data for the one-color scheme with $E_{\rm pump} = E_{\rm probe} = 1.503$~eV shown in Figure~\ref{Fig:3S} with the data recorded in the two-color scheme with $E_{\rm probe}=1.503$~eV and  $E_{\rm pump}=2.237$~eV ($\Delta=0.734$~eV) given in Figure~\ref{Fig:4S}. Note, that in both experiments the probe energy is the same.  The extracted $g$-factors and spin dephasing times, including their magnetic field and temperature dependences, are the same for one-color and two-color experiments. This is an expected result, as the carrier thermalization to the band gap occurs fast within a few picoseconds, and the long-lived coherent spin dynamics lasting up to a few nanoseconds are provided by localized electrons and holes. This is confirmed by the results shown in Figures~\ref{Fig:2}c,d, where the independence of the electron and hole spin dephasing times and their Larmor precession frequencies from the pump detuning are demonstrated. 


The dependences of the electron and hole KE amplitude on the pump detuning are shown in Figure~\ref{Fig:2}a. The electron amplitude decreases by about 30\% for the detuning increased from 0 to 0.05~eV and then stays constant up to $\Delta=0.43$~eV. For larger detunings it decreases and reaches zero for $\Delta=0.75$~eV. The hole KE amplitude shows a similar dependence, except of the absence of the initial decrease.

It is remarkable, that carrier spin coherence can be efficiently generated for large detunings and remains after carrier energy relaxation via emission of many optical and acoustic phonons. This evidences that the electrons and holes in the  studied crystal maintain their spin during energy relaxation and that the Dyakonov-Perel spin relaxation mechanism is absent in this crystal with spatial inversion symmetry~\cite{Kepenekian2015,Kepenekian2017,XOO2024_all}.  The dependence of the KE amplitude on detuning shown in Figure~\ref{Fig:2}a is in good agreement with our recent results including the detuning dependence of the optical orientation of excitons and carriers in the same material~\cite{XOO2024,COO2024}, which has also been found to be a common property of various lead halide perovskite crystals~\cite{XOO2024_all}. To illustrate that, we reproduce in Figure~\ref{Fig:2}b the detuning dependence of the optical orientation degree, measured under continuous-wave excitation, from ref.~\onlinecite{COO2024}. Our model consideration of the spin generation efficiency in ref.~\onlinecite{XOO2024} highlights that the decrease of the  optical orientation at detunings exceeding 0.3~eV is mainly provided by the deviation of the selection rules from being strict circular for optical transitions away from the R-point at higher momentum in the Brillouin zone. Then circularly polarized photons do not induce 100\% spin polarized carriers and excitons. Also, additional depolarization originates from spin-flip scattering of excitons during the energy relaxation, provided by the Elliott-Yafet mechanism.

\subsection{Mechanisms of generating spin coherence}
\label{mechanims}

Let us consider the mechanisms of optical generation of carrier spin coherence. In the experiment we observe coherent spin precession of electrons and holes on a time scale of few nanoseconds, which considerably exceeds the recombination time of excitons of about 0.1 ns.

One of the mechanisms is related to dissociation of the spin-oriented exciton into an electron and a hole without loosing their spin polarization. Then the carriers relax in energy and become localized in the vicinity of the band gap in spatially separated sites, so that their recombination is prohibited. When the spin orientation of carriers does not get lost (or gets only partially lost) during the energy relaxation, one can detect coherent spin precession of these carriers in an applied magnetic field. This mechanism requires pump energies exceeding the band gap energy.

Another mechanism suggests spin orientation of the resident carriers by means of spin-oriented excitons. This mechanism that exploits charged exciton complexes (trions) was suggested for semiconductor quantum wells hosting an electron gas of low density and for ensembles of singly charged quantum dots~\cite{Zhukov2007,Yakovlev_ch6_2017}. In simple terms, it can be explained as follows. The ensemble of resident carrier spins is initially unpolarized, so that their average spin projection on the light propagation direction is zero. Circularly polarized photons (say, with $\sigma^+$ polarization) generate excitons, whose spin is oriented along the light direction. Such an exciton can capture a resident carrier to form a trion in a spin singlet state, e.g., for the negatively charged trion it consists of two electrons with opposite spin orientation and a spin-polarized hole. As the exciton spin orientation is given by the light helicity the trion formation process is spin dependent, and a resident carrier with suitable spin polarization is taken from the ensemble for the formation process. As a result, the spin ensemble of resident carriers becomes polarized. This long-living polarization is maintained after the trion recombination, as the carrier that is returned to the ensemble after trion recombination does not necessarily still have its initial orientation. This orientation can get lost if spin relaxation in the trion occurs and/or in an external magnetic field due to differences in the Larmor precession frequencies of the trions and resident carriers. This mechanism explains the generation of spin coherence for resonant excitation of exciton and trion states at pump photon energies smaller than the band gap energy. It also can be effective for an exciton generated with large excess energy, when the exciton relaxes as a whole without dissociation and forms a trion with a resident carrier.

In the experiments on perovskite crystals using various pump detunings (Figure~\ref{Fig:2}a) both mechanisms are relevant. In our recent study of time-resolved photoluminescence in magnetic field on FA$_{0.9}$Cs$_{0.1}$PbI$_{2.8}$Br$_{0.2}$ crystals, we observed exciton spin beats in linear polarization, which were excited by a circularly polarized laser with an excess energy of 170~meV~\cite{XOO2024}. This evidences that a considerable fraction of excitons  does not dissociate after photogeneration, but relaxes as a whole. The relative contributions of the two mechanisms should vary with the pump detuning as well as depend on parameters such as lattice temperature, pump excitation density, etc.

\subsection{Optical control of carrier spin coherence}
\label{3 beams add}

Next, we turn to the problem of optical control of the carrier spin coherence and examine to what extent the established experimental approaches can be used for the lead halide perovskite crystals, namely for the FA$_{0.9}$Cs$_{0.1}$PbI$_{2.8}$Br$_{0.2}$ crystal studied here. The experimental realization of optical spin coherence control requires two pump pulses with the possibility to tune the delay between them and individually adjust their photon energy, intensity, and helicity. For conventional semiconductor nanostructures, such experiments were realized for the spins of Mn$^{+2}$ ions in CdTe/(Cd,Mn)Te quantum wells (QWs)~\cite{Akimoto1998}, and for the resident electrons in CdTe/(Cd,Mg)Te QWs~\cite{Zhukov2010}, in (In,Ga)As/GaAs quantum dots~\cite{Greilich2009,Phelps2009}, and in CdSe nanocrystals~\cite{Zhang2010}. In experiments with ensembles of spins two regimes can be realized.

For the additive regime, the second pump does not change the spin coherence of carriers induced by the first pump, but generates further spin coherence involving other carriers. The spin polarizations generated by the two pumps can either amplify or dampen each other, depending on the relation between phases of their Larmor precession motion, e.g. whether they are in phase or in antiphase with each other. The additive regime was realized in  refs.~\onlinecite{Akimoto1998,Zhukov2010}, where the same photon energies for both pump pulses were used.

For the spin control regime, one avoids the additional generation of spin polarization by the second pump, as in this case the goal is to use the second pump for optical rotation of the spin coherence generated by the first pump. For that, the photon energy of the second pump is detuned to a lower energy relative to the first pump, where the light absorption can be neglected~\cite{Greilich2009,Phelps2009}. However, the detuning should not be too large, as the control efficiency reduces strongly for larger detuning.  The spin control regime was demonstrated so far only for charge carriers strongly confined in (In,Ga)As quantum dots~\cite{Greilich2009}, CdSe nanocrystals~\cite{Zhang2010}, and recently for CsPbBr$_3$ nanocrystals~\cite{Lin2023}.

\begin{figure}[hbt!]
\includegraphics[width=8cm]{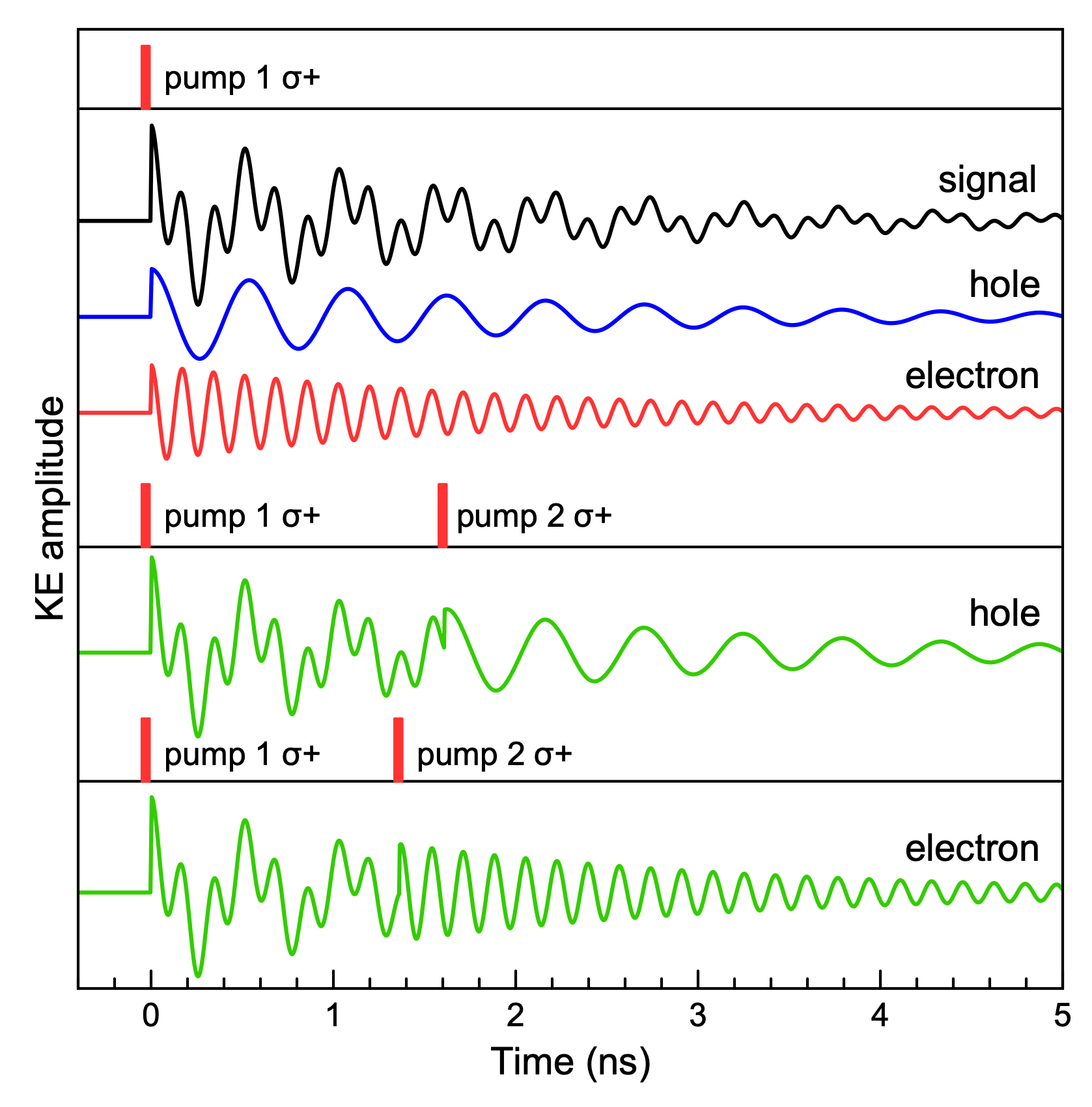}
\caption{Modeling of spin coherence control in the additive regime. The hole (blue line) and electron (red line) contributions are shown separately. Black line is their sum simulating the experimental signal with only one pump applied (pump 1). The model parameters are given in the text. Green lines are the spin dynamics provided by two pump pulses delayed by 1.61~ns or 1.36~ns relative to each other, so that either the electron or the hole components are selectively suppressed.  Red dashes indicate the arrival times of the circularly-polarized pump pulses.}
\label{Fig:3}
\end{figure}

Perovskite semiconductors offer interesting possibilities for investigating the optical spin control feasibility, as in them electron and hole spin coherence  coexist, see Figure~\ref{Fig:1}b. Therefore, it is attractive to explore the opportunity of individual control of electron and hole spins by enhancing or suppressing their contributions.  To illustrate this idea for the additive regime, we show in Figure~\ref{Fig:3} model calculations of the spin dynamics of resident electrons and holes induced by two pump pulses with the same photon energies ($E_{P1}=E_{P2}=E_{\rm probe}$) and the same circular polarizations ($\sigma^+$). For modeling we assume that the first pump generates the spin coherence of electrons and holes with the same efficiency: $A_{P1}^e=A_{P1}^h=1$. To achieve full suppression of either the hole or the electron contribution, the amplitude of the second pump was about 50\% of that of the first pump, namely $A_{P2}^e=0.5$ and $A_{P2}^h=0.5$. The used carrier $g$-factors and their spin dephasing times were set close to those measured for FA$_{0.9}$Cs$_{0.1}$PbI$_{2.8}$Br$_{0.2}$ crystals: $g_e=3.5$, $g_h=-1.1$, and $T_{2,e}^*=T_{2,h}^*=2$~ns.

In the upper part of Figure~\ref{Fig:3} the calculated spin dynamics induced by only one pump (pump 1) is shown by the black line. It consists of the hole (blue line) and the electron (red line) contributions. The green dynamics traces are calculated for the action of the two pump pulses. When their delay is set to 1.61~ns, pump 2 generates electron spin coherence with phase opposite to the electron coherence induced by pump 1, note the minimum amplitude of the red dynamics. At the same time this setting coincides with the maximum amplitude for the hole dynamics and results in amplification of the hole signal. It is clearly seen from the corresponding green dynamics that after pump 2 the electron contribution is fully suppressed and only the enhanced hole component is left. 

By setting the delay between pumps to 1.36~ns one can suppress the hole component and enhance the electron contribution, see the bottom dynamics in Figure~\ref{Fig:3}. Note that this approach allows one to "purify" spin signals when experimental studies of either the electrons or the holes need to be performed without decomposing their contributions.  For example in optically-detected NMR studies~\cite{kirstein2022am} or, in experiments without time resolution, where carrier contributions cannot be separated via their Larmor precession difference.

\begin{figure}[hbt!]
\includegraphics[width=8.75cm]{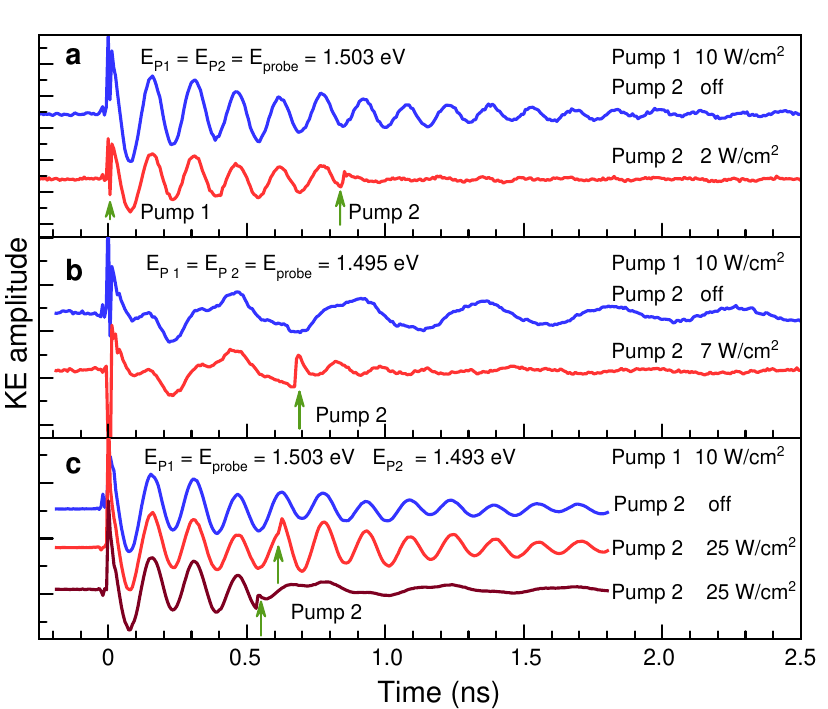}
\caption{Control of spin coherence in the FA$_{0.9}$Cs$_{0.1}$PbI$_{2.8}$Br$_{0.2}$ crystals in the additive regime. KE dynamics in two-pump experiments with different delays between the pumps and various pump powers.  (a) $E_{P1}=E_{P2}=E_{\rm probe}=1.503$~eV, pump 2 off (blue line) and $P_2=2$~W/cm$^{2}$ (red line). (b) $E_{P1}=E_{P2}=E_{\rm probe}=1.495$~eV, pump 2 off (blue line) and $P_2=7$~W/cm$^{2}$ (red line). (c) $E_{P1}=E_{\rm probe}=1.503$~eV, $E_{P2}=1.493$~eV, pump 2 off (blue line) and $P_2=25$~W/cm$^{2}$ (red and brown lines with different pump 2 arrival times). In all panels $P_1=10$~W/cm$^{2}$. Zero time delay is set to the moment of pump 1 pulse arrival. The delay of pump 2 is marked by the green arrows. $B_{\rm V}=0.125$~T and $T=1.6$~K.}
\label{Fig:4}
\end{figure}

An experimental realization of additive control is illustrated in Figure~\ref{Fig:4}. In order to measure the spin coherence generated by pump 1 and pump 2, the helicity of both pumps is simultaneously varied by a photoelastic modulator at the frequency of 50~kHz. The intensity of the probe is modulated at the frequency of~84 kHz and the signal is detected with a lock-in amplifier at the difference frequency, for details see the Supporting Information S2. The pump 1 power is  kept at $P_1=10$~W/cm$^2$ and the pump 2 power is varied from zero (pump 2 off) up to $P_2=25$~W/cm$^2$. By varying the pump photon energy we are able to tune the ratio between the electron and hole contributions, as they have different spectral dependences, see Figure~\ref{Fig:1}b. As one can see from the blue dynamics in Figures~\ref{Fig:4}a,c, measured for the pump 1 only with $E_{P1}=1.503$~eV, the KE dynamics is dominated by the electron contribution, while for $E_{P1}=1.495$~eV the electron and hole amplitudes are about equal to each other. 

Figure~\ref{Fig:4}a (red dynamics) shows that spin oscillations can be suppressed by employing pump 2, in which delay is adjusted to induce spin coherence with a phase opposite to that of the electron dynamics generated by pump 1. In Figure~\ref{Fig:4}b (red dynamics) the pump 2 delay is adjusted to suppress the hole contribution and enhance the electron signal. Indeed, at delays after pump 2 arrival only amplified electron precession is detected. These examples demonstrate that the concept of individual control of electron and hole spin coherence, which we have presented in Figure~\ref{Fig:3}, is working well for perovskite crystals. Figure~\ref{Fig:4}c shows that the two-color two-pump experimental protocol can also be implemented to control coherent spin precession. Here, $E_{P1}=1.503$~eV and pump 2 is detuned to lower energy at $E_{P2}=1.493$~eV. By varying the delay, the electron oscillations can be either enhanced (red dynamics) or suppressed (brown dynamics). More examples of spin control in the additive regime can be found in the Supporting Information Figure~S7.

\begin{figure}[hbt!]
\includegraphics[width=8.75cm]{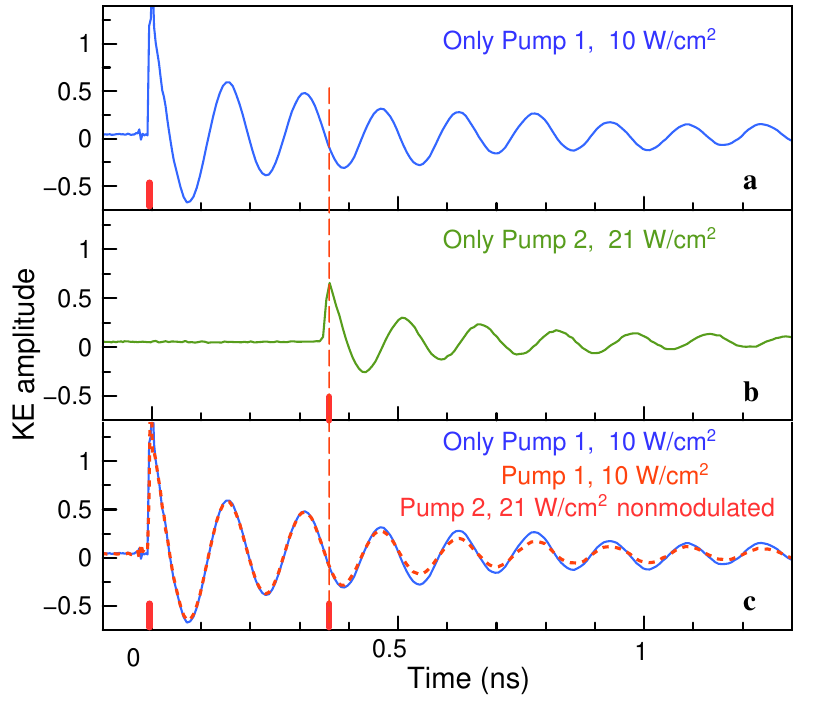}
\caption{Spin dynamics in the FA$_{0.9}$Cs$_{0.1}$PbI$_{2.8}$Br$_{0.2}$ crystals in the spin control regime. Here, the KE dynamics in the two-pump regime is given with $E_{P1}=E_{P2}=E_{\rm probe}=1.503$~eV.  $B_{\rm V}=0.125$~T and $T=1.6$~K. (a) KE dynamics induced by pump 1 only with $P_1=10$~W/cm$^2$. (b) KE dynamics induced by pump 2 only with $P_2=21$~W/cm$^2$. Pump 2 has $\sigma^+$ circular polarization and is intensity-modulated by a chopper at the frequency of 2~kHz. The signal is detected at the same frequency. (c) KE dynamics with pump 1 only (solid blue line) and with both pumps (dashed red line). Pump 2 has $\sigma^+$ circular polarization and is not modulated, so that its additive contribution to the signal is not detected. Red bars mark the arrival times of the pump pulses. In panels (a,c) the signals are detected using a lock-in amplifier at the frequency of 50~kHz.}
\label{Fig:5}
\end{figure}

We check also the feasibility of all-optical spin manipulation by means of the two-pump technique in the spin control regime. Several measurements are performed at different magnetic fields (from 0.125~T up to 0.5~T), at different energy detunings between the pumps (from $\Delta=0$ up to 5~meV), and at various time delays between the pumps. The experimental traces are practically the same for all these conditions. They are illustrated in Figure~\ref{Fig:5} for $\Delta=0$~meV, $B_{\rm V}=0.125$~T, and the delay of 0.36~ns.  Blue lines in Figures~\ref{Fig:5}a,c give the spin dynamics induced by the pump 1 only, which helicity is modulated between $\sigma^+$ and $\sigma^-$  at the frequency of 50~kHz. At the selected photon energy ($E_{P1}=1.503$~eV) the signal is mainly contributed by the electron spin precession.

In order to demonstrate the impact of the delay of pump 2, we show in Figure~\ref{Fig:5}b the spin dynamics induced solely by the pump 2 action. For that, the pump 2 intensity is modulated at the frequency of 2~kHz, the detection is done at the same frequency, and the polarization is kept constant at $\sigma^+$. Here the electron spin precession starting at 0.36~ns delay is clearly seen. Note that the delay is set to the moment, at which the spin dynamics amplitude induced by pump 1 (blue line) crosses zero. In this case, the spin polarizations induced by pump 1 and pump 2 are perpendicular to each other. For this setting, rotation of the spin polarization induced by the pump 1 is expected, showing up as a phase shift~\cite{Greilich2009,Lin2023}.

The dashed red line in Figure~\ref{Fig:5}c shows the dynamics after action of both pumps. We are interested in the action of pump 2 on the spin coherence generated by pump 1. Therefore, in order to exclude a contribution of spin coherence directly generated by pump 2, we use here a nonmodulated pump 2, which contribution is screened by the lock-in detection. Note that the action of pump 2 at $\pm\pi$ phase of the carrier precession, due to the $\sigma^\pm$ modulation of pump 1, is similar, see the Bloch sphere concept in ref.~\onlinecite{Greilich2009}. By comparing the solid blue and dashed red dynamics one can conclude that no phase shift takes place due to the action of pump 2. Only a faster decay of the signal amplitude is induced by pump 2. The decay time changes from 700~ps for the blue dynamics to 520~ps for the red dynamics. More details for other magnetic fields and pump powers are given in the Supporting Information Figure~S8. No phase shift is found in these cases, but an increase of the pump 2 power results in a stronger decrease and a faster decay of the spin coherence amplitude.

To summarize, we are not able to achieve optical spin rotation for carrier spins  in the studied FA$_{0.9}$Cs$_{0.1}$PbI$_{2.8}$Br$_{0.2}$ crystal. We suggest that the main reason is the rather weak localization of the resident electrons and holes. We remind that so far optical spin rotation has been realized only for electrons strongly confined in quantum dots or nanocrystals. The effect of pump 2 on the signal amplitude can be explained by heating of the resident carriers and their delocalization. This is in line with our previous findings of a strong dependence of the carrier spin dynamics in FA$_{0.9}$Cs$_{0.1}$PbI$_{2.8}$Br$_{0.2}$ crystals on the lattice temperature~\cite{kirstein2022am}. The spin dephasing time shortens considerably and the KE amplitude is reduced for a temperature increase from 1.6~K up to 25~K, see the Supporting Information Figure~S4.

\section{Conclusions}

The coherent spin dynamics of electrons and holes in a FA$_{0.9}$Cs$_{0.1}$PbI$_{2.8}$Br$_{0.2}$ perovskite bulk crystal have been studied experimentally using one-color and two-color pump-probe schemes. In the Kerr ellipticity signal, Larmor precession of electron and hole spins is observed in a magnetic field, even when the detuning between the photon energies of the pump and probe reaches 0.75~eV. This evidences that the spin polarization of electrons and holes is preserved during their fast energy relaxation, in line with recent results on optical orientation of excitons and carriers in this material~\cite{XOO2024,COO2024}. By using two-pump protocols we demonstrate that the spin coherence of resident electrons and holes can be controlled in the additive regime. Namely, their amplitudes can be reduced or enhanced by tuning the delay and power of the second pump. The regime of all-optical spin control has not been achieved due to the relatively weak localization of the resident carriers in the studied bulk crystals. The demonstration of spin control in the additive regime is, however, an important technique allowing to purify the spin signal towards a single component, especially helpful for perovskite semiconductors. These results confirm that the lead halide perovskite semiconductors are  promising and suitable materials for spintronics and spin-optronics applications.



\section{Methods}

\textbf{Samples.}
The studied {FA}$_{0.9}$Cs$_{0.1}$PbI$_{2.8}$Br$_{0.2}$ single crystal was grown out of solution using the inverse temperature crystallization technique~\cite{nazarenko2017}, see also the Supporting Information S1. Details of the optical characterization of this sample and measurements of the coherent spin dynamics of electrons and holes under resonant excitation can be found in ref.~\onlinecite{kirstein2022am}.

\textbf{Magneto-optical measurements.} The sample was placed in an optical magneto-cryostat (containing a pair of split coils) with the temperatures variable from 1.6~K up to 30~K. For $T=1.6$~K, the sample was immersed in superfluid helium, while for $T=4.2–30$~K the sample was hold in cooling helium gas. The measurements were performed in magnetic fields up to 1.5~T oriented perpendicular to the direction of the laser pump beam (Voigt geometry, $B_{\rm V}$). For our measurements of the coherent spin dynamics and population dynamics of carriers we used a time-resolved pump-probe technique in various schemes, using detection of the Kerr ellipticity or the differential reflectivity. Details of these methods are given in the Supporting Information S2.

\textbf{ASSOCIATED CONTENT}

\textbf{Supporting Information.}

Additional information on the sample synthesis, experimental techniques, population dynamics measured with time-resolved differential reflectivity, and details of the spin dynamics in the one-color and and two-color schemes and of the two-color two-pump pulse spin control regime.

\textbf{AUTHOR INFORMATION}

{\bf Corresponding Authors} \\
Evgeny~A.~Zhukov, Email: evgeny.zhukov@tu-dortmund.de\\
Dmitri R. Yakovlev,  Email: dmitri.yakovlev@tu-dortmund.de\\

\textbf{ORCID}\\
Evgeny~A.~Zhukov:     0000-0003-0695-0093   \\
Dmitri R. Yakovlev:   0000-0001-7349-2745\\
Erik~Kirstein:        0000-0002-2549-2115 \\
Nataliia E. Kopteva:   0000-0003-0865-0393 \\
Oleh Hordiichuk:      0000-0001-7679-4423 \\
Maksym V. Kovalenko:  0000-0002-6396-8938\\
Manfred Bayer:        0000-0002-0893-5949 \\




{\bf Notes}\\
The authors declare no competing financial interests.

{\bf Acknowledgments}\\
The authors are thankful to A. Greilich, I. A. Akimov, and  M. M. Glazov for fruitful discussions.
We acknowledge the financial support by the Deutsche Forschungsgemeinschaft via the SPP2196 Priority Program (project YA 65/28-1, no. 527080192). N.E.K. acknowledges support of the Deutsche Forschungsgemeinschaft (project KO 7298/1-1, no. 552699366). The work at ETH Z\"urich (O.H., D.N.D., and M.V.K.) was financially supported by the Swiss National Science Foundation (grant agreement 200020E 217589, funded through the DFG-SNSF bilateral program) and by the ETH Z\"urich through the ETH+ Project SynMatLab.


\onecolumngrid
\vspace{\columnsep}
\begin{center}
\newpage
\makeatletter
{\large\bf{Supporting Information\\ Coherent spin dynamics of electrons and holes photogenerated with large kinetic energy in lead halide perovskite crystals}}\\

Evgeny~A.~Zhukov, Dmitri~R.~Yakovlev, Erik~Kirstein, Nataliia~E.~Kopteva, Oleh Hordiichuk, Maksym V. Kovalenko, and Manfred~Bayer
\makeatother
\end{center}
\vspace{\columnsep}


\twocolumngrid

\setcounter{page}{1}
\setcounter{section}{0}
\setcounter{equation}{0}
\setcounter{figure}{0}

\renewcommand{\thesection}{S\arabic{section}}
\renewcommand{\thepage}{S\arabic{page}}
\renewcommand{\theequation}{S\arabic{equation}}
\renewcommand{\thefigure}{S\arabic{figure}}
\renewcommand{\bibnumfmt}[1]{[S#1]}
\renewcommand{\citenumfont}[1]{S#1}


\section{Material information}
\label{Sample}

The studied perovskite crystals are based on the FAPbI$_{3}$ material class. FA-based perovskite exhibits a low trap density ($1.13 \times 10^{10}$~cm$^{-3}$) and a low dark carrier density ($3.9 \times 10^{9}$~cm$^{-3}$)~\cite{nazarenko2017si,zhumekenov2016si}. FAPbI$_3$ is chemically and thermally more stable compared to MAPbI$_3$ due to the decomposition of the latter to gaseous hydrogen iodide and methylammonium~\cite{nazarenko2017si}. However, pure FAPbI$_3$ suffers from structural instability originating from the large size of the FA cation, which cannot be accommodated by the inorganic perovskite framework. This instability has been successfully resolved via partial, up to 15\%, replacement of the large FA cation with smaller caesium (Cs) together with iodine (I) substitution by bromide (Br) \cite{mcmeekin2016_SI,jeon2015_SI}. As a result, the Goldschmidt tolerance factor~\cite{goldschmidt_gesetze_1926si} $t$ is tuned from 1.07 in FAPbI$_3$ closer to 1 in FA$_{0.9}$Cs$_{0.1}$PbI$_{2.8}$Br$_{0.2}$, where it is 0.98. The band gap of FA$_{0.9}$Cs$_{0.1}$PbI$_{2.8}$Br$_{0.2}$ at room temperature is 1.52~eV, slightly larger than the band gap of FAPbI$_{3}$ of 1.42~eV~\cite{li2016si,nazarenko2017si}.

For crystal synthesis the inverse temperature crystallization technique is used~\cite{nazarenko2017si,zhumekenov2016si}. For the growth, a solution of CsI, FAI (FA being formamidinium), PbI$_2$, and PbBr$_2$, with GBL $\gamma$-butyrolactone as solvent is mixed. This solution is then filtered and slowly heated to 130$^\circ$C temperature, whereby the single crystals are formed in the black phase of FA$_{0.9}$Cs$_{0.1}$PbI$_{2.8}$Br$_{0.2}$, following the reaction
\begin{equation*}
\mathrm{PbI}_2 + \mathrm{PbBr}_2 + \mathrm{FAI} + \mathrm{CsI} \underset{}{\stackrel{[GBL]}{\rightarrow}} \mathrm{FA}_{0.9}\mathrm{Cs}_{0.1}\mathrm{PbI}_{2.8}\mathrm{Br}_{0.2} + \mathrm{R}.
\end{equation*}
Afterwards the crystals are separated by filtering and drying. A typical crystal used for this study  has a size of about 2~mm. The crystallographic analysis suggests that one of the principal axis $a$, $b$, $c$ is normal to the front facet, thus pointing along the optical axis. In cubic approximation, $a=b=c$. The pseudo-cubic lattice constant for hybrid organic perovskite (HOP) is around 0.63~nm~\cite{whitfield2016si}, but was not determined for this specific sample.

\section{Experimental techniques}

\textbf{Photoluminescence and optical transmission.}

Photoluminescence (PL) spectra are measured under continuous-wave (cw) diode laser excitation with 3.06~eV photon energy (wavelength of 405~nm). The PL emission is detected with a Si-based charge-coupled-device camera attached to a 0.5~m spectrometer.

\textbf{Time-resolved Kerr ellipticity.}

To study the coherent spin dynamics of carriers we used a time-resolved pump-probe technique with detection of the Kerr ellipticity (KE). This technique was successfully used to study both semiconductor structures (see, for example, refs.~\cite{Awschalom2002si,Yakovlev_ch6_2017si}) and various perovskites (see refs.~\cite{belykh2019si,kirstein2022amsi,kirstein2022ncsi,odenthal2017si,garcia-arellano2021si,Crane2020si,Grigoryev2021si}). Spin coherence of electrons and holes was generated by circular-polarized pump pulses (duration 1.5~ps, spectral width about 1~meV) emitted by a Ti:Sapphire mode-locked laser operating at a repetition frequency of 75.8~MHz (repetition period $T_R=13.2$~ns). Carrier spin polarization is generated along the light vector $\mathbf{k}$.
The pump beam helicity was modulated between $\sigma^+$ and $\sigma^-$ polarizations at the frequency of 50~kHz using a photoelastic modulator (PEM). The pump-excited area of the sample (about 300~$\mu$m in diameter) was monitored by linearly-polarized probe pulses in the reflection geometry. The probe beam area is slightly smaller than that of the pump beam. The pump power density was tuned in the range $P_\text{pump}=0.4-70$~W/cm$^2$ and the probe power density was $P_\text{probe}=1.2-2.2$~W/cm$^2$. The induced ellipticity of the probe pulses was measured as function of the delay between the pump and the probe pulses using a balanced detector connected to a lock-in amplifier. Here, a double modulation scheme was used. The intensity of the probe beam was modulated with 84~kHz  frequency and the signal was detected at the difference frequency of the pump and probe modulation.

Magnetic fields up to 3~T are applied perpendicular to the pump laser beam wave vector, $\textbf{B}_V \perp \textbf{k}$ (Voigt geometry). The experiments are performed at cryogenic temperatures in the range $1.6-30$~K.

We used several measurement schemes.

(a) \textbf{One-color scheme.} Here, the photon energies of the pump and probe are the same, also when they are tuned.

(b) \textbf{Two-color scheme}.  This scheme is possible by using two Ti:Sapphire mode-locked lasers synchronized in time with each other with an accuracy of 100~fs. One laser was used as pump, and the other one as probe. To expand the tuning range of the pump laser for $1.796-2.361$~eV we use an optical parametric oscillator.

(c) \textbf{Two-color two-pump scheme}. To study the possibility of ultrafast optical rotation of spins we use one pump beam and the probe beam with the same photon energy and a second pump beam with a different energy. Also the time delay between the pumps is tuned. The polarizations of both pumps can be synchronously modulated by the same modulator between $\sigma^+$ and $\sigma^-$ (additive regime~\cite{Zhukov2010si}), or only the first pump is modulated (spin control regime~\cite{Greilich2009si,Burkard2000si,Pryor2006si,Lin2023si}). In the second case, the second pump is polarized either $\sigma^+$ or $\sigma^-$.

(d) \textbf{Time-resolved differential reflectivity.} Here, a fraction of the probe beam incident on the sample is directed to one channel of a balanced photodiode, and the probe beam reflected from the sample is directed to the second channel. When the pump and probe beams are linearly polarized, the time-resolved differential reflectivity TRDR ($\Delta R/ R$), which is determined by the population dynamics of the excited state, is measured. The measured signal is fitted using the following equation:
\begin{equation}
\dfrac{\Delta R}{R}=\sum_{i}{\left(\dfrac{\Delta R}{R}\right)_{0,i}\exp\left(-\frac{t}{\tau_{i}}\right)}.
 \label{n3}
\end{equation}
Here $(\Delta R/R)_{0,i}$ are the amplitude of the components of signal, $\tau_{i}$ are the population decay times, $i=1,2$. In these measurements the pump and probe beams have the same photon energy. 


\section{Population dynamics measured by differential reflectivity}
\label{TRGR}

\begin{figure}[hbt]
\includegraphics[width=8.75cm]{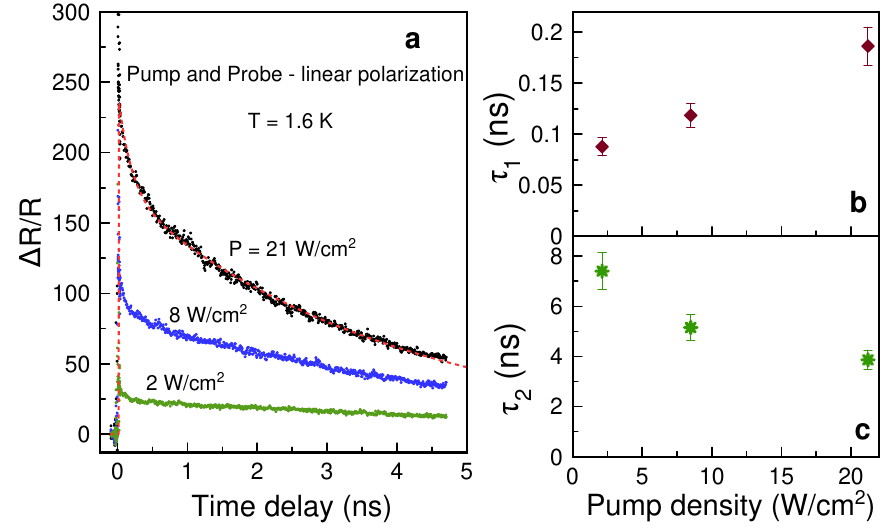}
\caption{Dynamics of differential reflectivity ($\Delta R/R$) in the FA$_{0.9}$Cs$_{0.1}$PbI$_{2.8}$Br$_{0.2}$ crystal. $E_{\rm pump}=E_{\rm probe}=1.504$~eV. Linearly polarized pump and probe beams.  (a) Dependence of $\triangle R/R$ on the time delay for different pump densities.  Fits with eq~\eqref{n3} are shown by the red lines. (b),(c) Dependence of $\tau_1$ and $\tau_2$ on pump density. $T=1.6$~K. }
\label{Fig:6S}
\end{figure}

We measure the population dynamics of excitons and carriers by means of  differential reflectivity. The dynamics of the $\Delta R/R$ signal at different pump powers measured for resonant exciton excitation is shown in Figure~\ref{Fig:6S}a. In these measurements, the pump and probe are linearly polarized. Fitting of these dynamics using eq~\eqref{n3} gives two components: a fast one with a decay time of about 150~ps and a slow one with a decay time of a few nanoseconds. The presence of such a long component is noteworthy. This component is likely due to localized (long-lived) carriers that are generated by the excitation, which we call resident carriers. The short-lived component is assigned to the exciton recombination dynamics. The short component decay time increases from 0.08~ns to 0.18~ns with increasing pump density, while the decay time of the second component decreases from 7~ns to 4~ns, see Figures~\ref{Fig:6S}b,c.

\section{Details of spin dynamics in the one-color regime}

\begin{figure*}[hbt]
\begin{center}
\includegraphics[width=16cm]{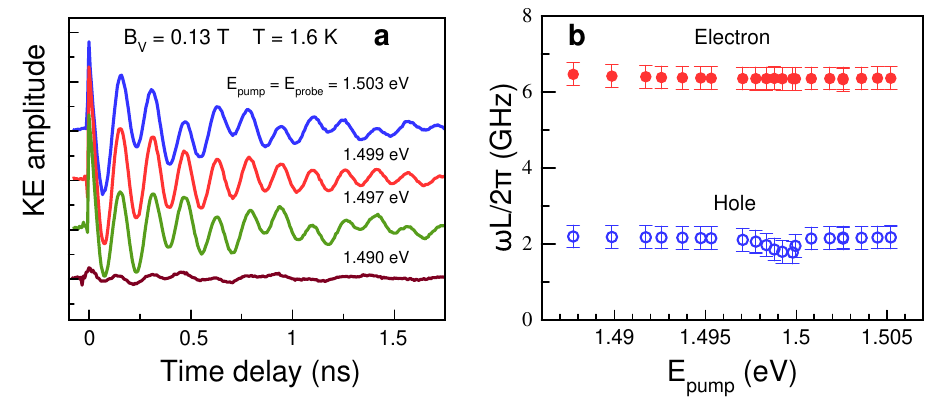}
\caption{Coherent spin dynamics in the FA$_{0.9}$Cs$_{0.1}$PbI$_{2.8}$Br$_{0.2}$ crystal measured using the one-color scheme at $B_{\rm V} = 0.13$~T and for $T=1.6$~K. (a) KE dynamics  at different energies $E_{\rm pump}=E_{\rm probe}$. (b) Spectral dependence of the electron and hole Larmor precession frequency.}
\label{Fig:1S}
\end{center}
\end{figure*}

\begin{figure*}[hbt]
\includegraphics[width=15cm]{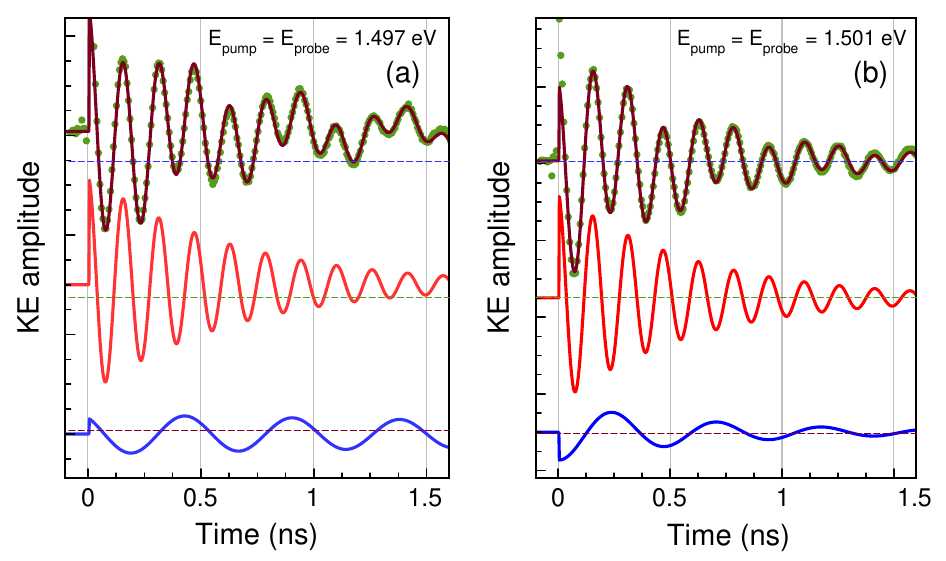}
\caption{Examples of fitting the KE dynamics measured in the one-color scheme for (a) $E_{\rm pump}=E_{\rm probe}=1.497$~eV and (b) $E_{\rm pump}=E_{\rm probe}=1.501$~eV. Experimental data are given by the green circles, brown line is a fit with eq~\eqref{n1}; also the individual components for electrons (red line) and holes (blue line) are shown. Note that the electron dynamics has the same phase in both panels, but the hole phase is opposite in the two panels.  Dashed lines show the zero signal levels. $P_{\rm pump}=7$~W/cm$^{2}$ and $P_{\rm probe}=1$~W/cm$^{2}$, $T=1.6$~K. 
}
\label{Fig:2S}
\end{figure*}

Detailed information on the measurements of the KE dynamics in one-color scheme at various photon energies are given in Figure~\ref{Fig:1S}. Here we present the KE dynamics and the spectral dependences of the Larmor precession frequencies for resident electrons and holes as well as their amplitude and phase. Part of these data is included in Figure 1 of the main text. Examples of fitting the KE dynamics at photon energies of 1.497~eV and 1.501~eV are given in Figure~\ref{Fig:2S}. A significant change in the phase of the signal is clearly observed.

\begin{figure*}[hbt]
\begin{center}
\includegraphics[width=17cm]{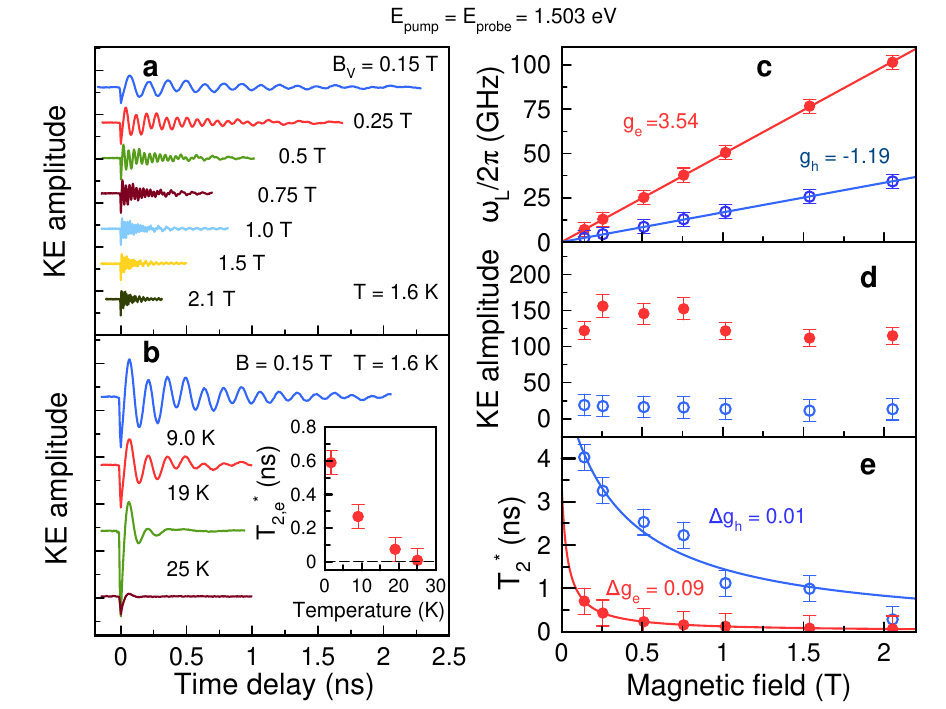}
\caption{Coherent spin dynamics in the FA$_{0.9}$Cs$_{0.1}$PbI$_{2.8}$Br$_{0.2}$ crystal measured in the one-color scheme with $E_{\rm pump}=E_{\rm probe}=1.503$~eV. $T=1.6$~K. (a) KE dynamics measured at different magnetic fields $B_{\rm V}$.  (b) KE signal at different temperatures. $B_{\rm V} = 0.15$~T. Inset shows the temperature dependence of $T_{2,e}^*$. (c) Magnetic field dependences of the Larmor precession frequency for electrons and holes. Linear fits to these data are shown by the lines. (d) Magnetic field dependences of the electron and hole KE amplitude.  (e) Magnetic field dependences of the electron and hole spin dephasing time $T_{2}^*$. Fits with eq~\eqref{n2} are shown by the lines. In panels (c), (d) and (e) $T=1.6$~K, the data for electrons are shown by the red solid circles and for holes by the blue open circles.   }
\label{Fig:3S}
\end{center}
\end{figure*}

Figure~\ref{Fig:3S} shows the evolution of the KE dynamics in the one-color scheme ($E_{\rm pump}=E_{\rm probe}=1.503$~eV) with increasing magnetic field from 0.15~T to 2.1~T and at different temperatures in the range from 1.6~K to 25~K, as well as the results of their fitting with eq~\eqref{n1}. With increasing magnetic field the oscillation frequency of the electron and hole components grows linearly (Figure~\ref{Fig:3S}c). From the linear dependence of the Larmor precession frequency on the magnetic field we evaluate the $g$-factors for electrons and holes using eq~\eqref{n2}: $g_{e}=3.54\pm0.11$ and $g_{h}=-1.19\pm0.05$. The amplitudes of the electron and hole components depend very weakly on the magnetic field (Figure~\ref{Fig:3S}d), but the spin dephasing times decrease with increasing field (Figure~\ref{Fig:3S}e). The dependences of $T_{2,e}^*$ and $T_{2,h}^*$ on  magnetic field are fitted with~\cite{Glazov2018si}:
\begin{equation}
\dfrac{1}{T_{2,e(h)}^*(B)}= \dfrac{1}{T_{2,0,e(h)}^*} + \dfrac{\Delta g_{e(h)}\mu_{\rm B}B}{\hbar},
\label{n4}
\end{equation}
where $\Delta g_{e(h)}$ is the dispersion of the electron (hole) $g$-factors, $T_{2,0,e(h)}^*$ is their spin dephasing times at zero magnetic field. The fit, which is shown by the red and blue lines for electrons and holes, respectively, in Figure~\ref{Fig:3S}e, gives the following parameters: $\Delta g_{e}=0.09$ ($\Delta g_{e}/g_{e}=2.5\%$), $\Delta g_{h}=0.01$ ($\Delta g_{h}/g_{h}=0.8\%$), $T_{2,0,e}^*=3.4$~ns, and $T_{2,0,h}^*=5.9$~ns.

The strong decrease of the KE amplitude with growing temperature and its disappearance for $T> 25$~K (Figure~\ref{Fig:3S}b), accompanied by shortening of the spin dephasing time (insert in Figure~\ref{Fig:3S}b), confirms the small localization energy of the resident electrons and holes contributing to the KE signals.

\clearpage

\section{Details of spin dynamics in the two-color scheme}

\begin{figure*}[hbt!]
\begin{center}
\includegraphics[width=15cm]{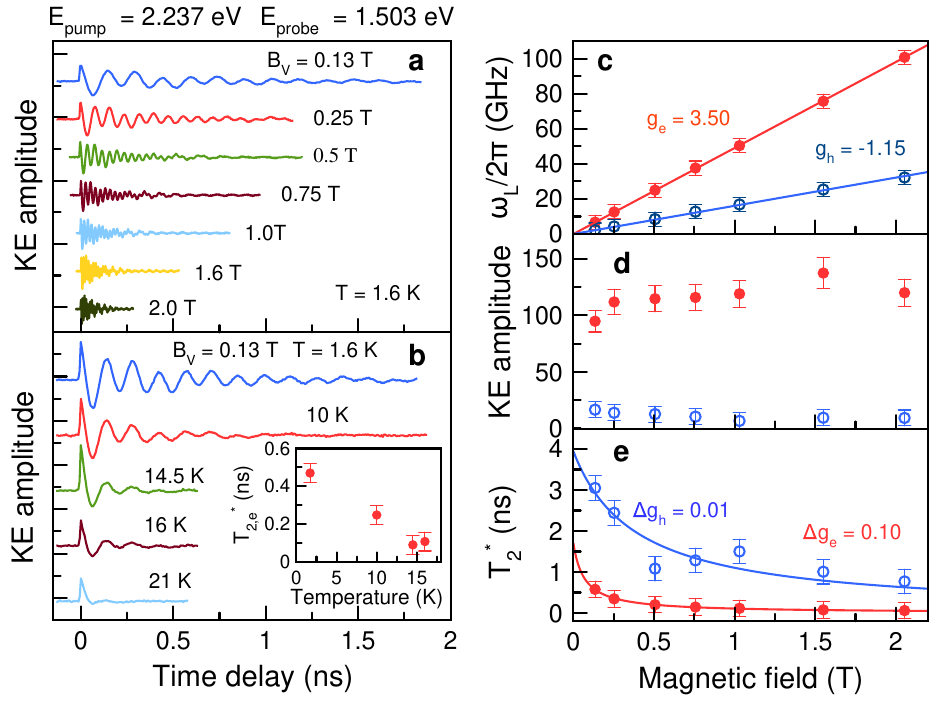}
\caption{Coherent spin dynamics in the FA$_{0.9}$Cs$_{0.1}$PbI$_{2.8}$Br$_{0.2}$ crystal measured in the two-color scheme with $E_{\rm pump}=2.237$~eV and $E_{\rm probe}=1.503$~eV. (a) KE dynamics measured in different magnetic fields at $T=1.6$~K. (b) KE dynamics measured at different temperatures in $B_{\rm V} = 0.13$~T. Insert shows the temperature dependence of $T_{2,e}^*$. (c) Magnetic field dependence of the Larmor precession frequency for electrons and holes. Lines give linear fits of the experimental data shown by circles. (d) Magnetic field dependence of the Kerr amplitudes of the electron and hole components. (e) Magnetic field dependence of spin dephasing times $T_{2}^*$ of electrons and holes. Fits with Eq.~\eqref{n2} are given by lines. In panels (c-e) $T=1.6$~K, the data for electrons are given by the red circles and for holes by the blue circles. 
}
\label{Fig:4S}
\end{center}
\end{figure*}

In Figures~\ref{Fig:4S}a,b we present data on the spin dynamics of carriers in different magnetic fields and temperatures measured using the two-color scheme with $E_{\rm pump}=2.237$~eV and $E_{\rm probe}=1.503$~eV.  Results of their fitting with eq~\eqref{n1} are given in Figures~\ref{Fig:4S}c-e.  From the linear dependence of the Larmor precession frequency on magnetic field we determine the $g$-factors for electrons and holes using eq~\eqref{n2}: $g_{e}=3.50\pm0.11$ and $g_{h}=-1.15\pm0.05$ (Figure~\ref{Fig:4S}c). The amplitudes of the electron and hole components weakly depend on magnetic field (Figure~\ref{Fig:4S}d), but the spin dephasing times decrease with increasing field (Figure~\ref{Fig:4S}e). The dependences of $T_{2,e}^*$ and $T_{2,h}^*$ on magnetic field fitted with eq~\eqref{n4} give the following parameters for electrons  $\Delta g_{e}=0.10$ ($\Delta g_{e}/g_{e}=2.8\%$) and $T_{2,0,e}^*=1.7$~ns, and for holes $\Delta g_{h}=0.01$ ($\Delta g_{h}/g_{h}=0.9\%$)  and $T_{2,0,h}^*=3.9$~ns.

\begin{figure*}[hbt!]
\includegraphics[width=14cm]{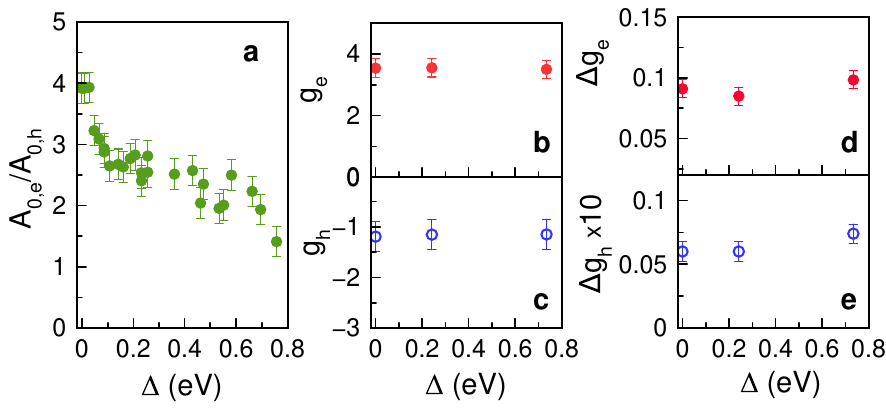}
\caption{Two-color scheme. Dependences on pump detuning measured at $E_{\rm probe}=1.503$~eV for: (a) the ratio of the electron and hole amplitudes in the KE signals, (b,c) the electron and hole $g$-factors, and (d,e) the dispersion of the electron and hole $g$-factors. $B_{\rm V} = 0.13$~T and $T=1.6$~K. }
\label{Fig:5S}
\end{figure*}

Figure~\ref{Fig:5S}a shows the dependence of the ratio of the electron to hole KE amplitudes on the pump detuning from $\Delta=0$~eV up to $0.735$~eV. These data are calculated from the data of Figure~\ref{Fig:2}a. The ratio shows a moderate initial increase, but then stays on the same level. The spread of $g$-factors of holes and electrons increases very little with increasing detuning (Figures~\ref{Fig:5S}d,e), while the $g$-factors do not change (Figures~\ref{Fig:5S}b,c).

\clearpage

\section{Two-color two-pumps measurements in spin control scheme}

\begin{figure*}[hbt]
\begin{center}
\includegraphics[width=16cm]{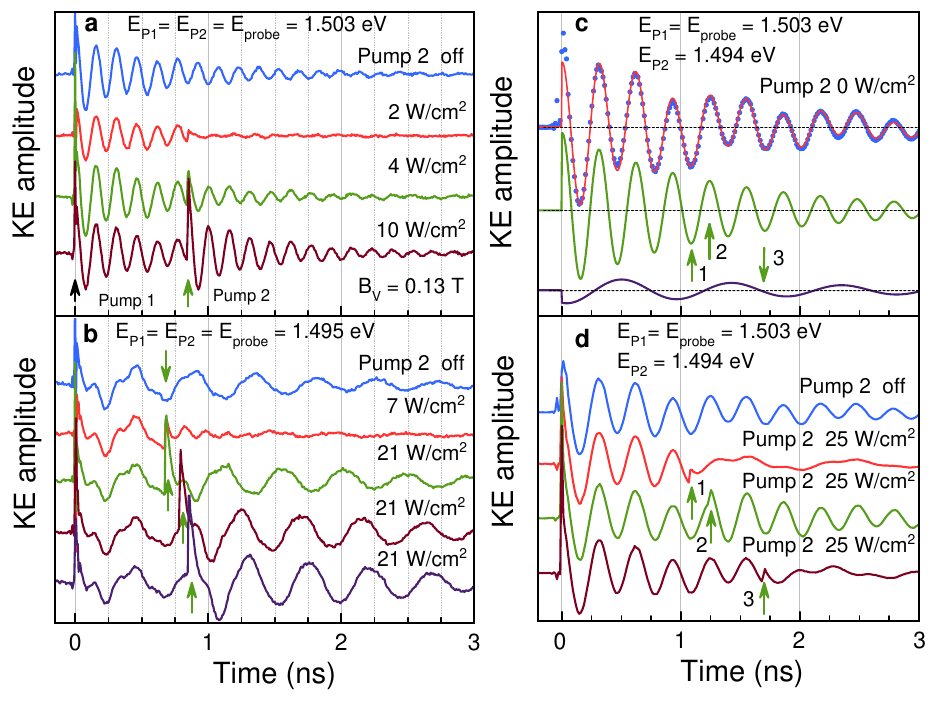}
\caption{KE signal for two-beam pumping in the additive regime. (a) KE dynamics at different powers of $P_2$, $E_{P1}=E_{P2}=E_{\rm probe}=1.503$~eV, identical delay between the two pumps. (b) KE dynamics at $E_{P1}=E_{P2}=E_{\rm probe}=1.495$~eV for different powers of $P_2$ and different delays between pumps. (c) KE signal (black dots) at $E_{P1}=E_{\rm probe}=1.503$~eV, $E_{P2}=1.494$~eV, pump 2 is switched off. Red line is fit with eq~\eqref{n1}. Green and blue lines give the electron and hole two components, respectively. (d) KE dynamics at $E_{P1}=E_{\rm probe}=1.503$~eV, $E_{P2}=1.494$~eV, different $P_2$ and different delays. The pump 1 arrival moment is set to time zero.  The pump 2 delay is marked with green arrows. Their positions are the same in (c) and (d). $P_1=10$~W/cm$^{2}$ for all curves. $P_{\rm probe}=1$~W/cm$^{2}$, $B_{\rm V}=0.125$~T, and $T=1.6$~K.}
\label{Fig:7S}
\end{center}
\end{figure*}

The results of measurements using the three-beam scheme in the additive mode (recall that in this case the two pumps are modulated analogously by a PEM at the same frequency) for various parameters of the pumps are shown in Figure~\ref{Fig:7S}. In all cases, $P_1=10$~W/cm$^{2}$ and $P_{\rm probe}=1$~W/cm$^{2}$. It should be noted that the result of the action of the pump 2 pulses on the electron spin polarization induced by pump 1 depends on two factors: (i) the intensity of pump 2, and (ii) the time delay between pump 1 and pump 2 (namely the phase of the carrier spin precession caused by pump 1 at the moment of pump 2 arrival). The time delay of the pump 2 pulse in Figure~\ref{Fig:7S} is marked with green arrows.

Figure~\ref{Fig:7S}a demonstrates results for the one-color scheme (zero energy detuning between pumps, $E_{P1}=E_{P2}=E_{\rm probe}=1.503$~eV) in the additive regime. At this energy, the KE dynamics is mainly contributed by the electron Larmor precession. The time delay of the pump 2 is set to $t_{12}=0.85$~ns, where the spin precession induced by pump 1 (blue line) has a minimum. Therefore, pump 2 generates carrier spin polarization in the opposite direction. For $P_2=2$~W/cm$^{2}$ (red line), the signal of pump 1 and pump 2 approximately compensate each other and no beats are seen after pump 2 arrival. With further increase of the pump 2 power, the signal amplitude with opposite phase is increasing at delays after pump 2 arrival.


The spin dynamics presented in Figure~\ref{Fig:7S}b are measured at $E_{P1}=E_{P2}=1.495$~eV, where the electron and hole contributions are about equal in strength and can be clearly seen in the blue dynamics resulting from the action of pump 1 only. For the red dynamics, the intensity ($P_2=7$~W/cm$^{2}$) and delay of pump 2 ($t_{12}=0.69$~ns are chosen such that the hole component is fully suppressed. With increasing power to $P_2=21$~W/cm$^{2}$ at the same delay, hole spin beats appear after pump 2 arrival with the opposite phase. By changing the delay of  pump 2 for the same power, one can see that the phase of the hole precession shifts with delay.

In Figures~\ref{Fig:7S}c,d the results of the two-color two-pump experiment ($E_{P1}=E_{\rm probe}=1.503$~eV and $E_{P2}=1.494$~eV) in the additive regime are shown. One can see that by adjusting the pump 2 delay the electron or hole components can be either suppressed or enhanced.

\begin{figure*}[hbt]
\begin{center}
\includegraphics[width=16cm]{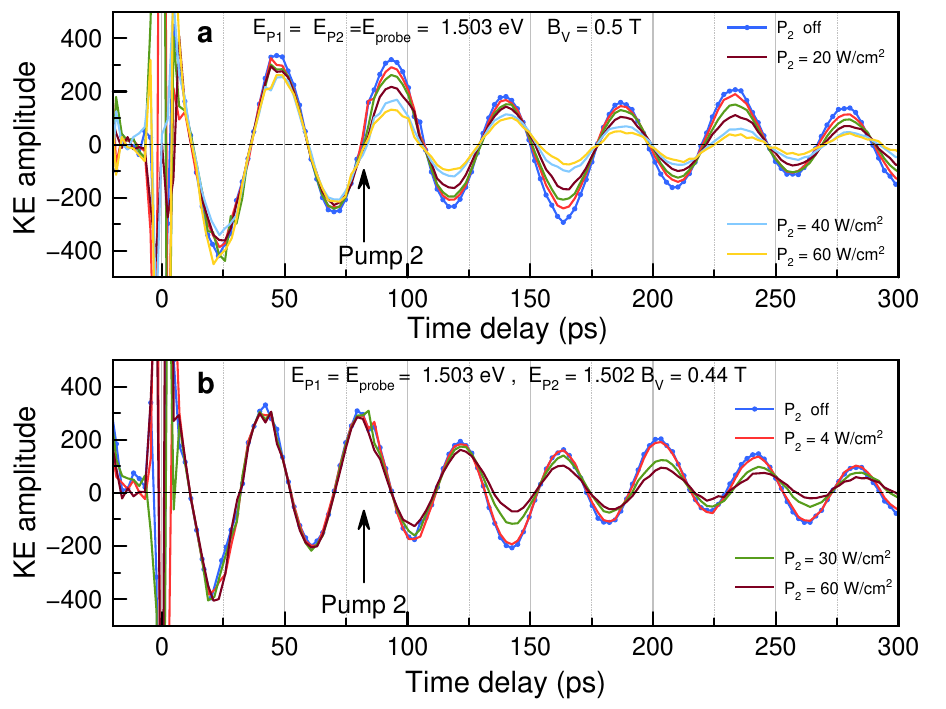}
\caption{KE dynamics in the two-pump experiment applying the spin control regime. (a) KE dynamics at $E_{P1}=E_{P2}=E_{\rm probe}=1.503$~eV, different powers of $P_2$ and the same delay between pumps. $B_{\rm V}=0.5$~T. (b) KE dynamics at $E_{P1}=E_{\rm probe}=1.503$~eV, $E_{P2}=1.502$~eV, different powers of $P_2$ and the same delay. $B_{\rm V}=0.44$~T.  The moment of the first pump pulse arrival corresponds to zero time delay. The pump 2 delay is marked with black arrows. $P_1=3$~W/cm$^{2}$ for all dynamics, $P_{\rm probe}=1$~W/cm$^{2}$, $T=1.6$~K.}
\label{Fig:8S}
\end{center}
\end{figure*}

KE dynamics in the two-pump experiment in the spin control regime at different detunings and different pump 2 powers are shown in Figure~\ref{Fig:8S}. One can see that the variation of detuning and pump 2 power do not lead to a change in the phase of the signal after pump 2 arrival, but only reduces the signal amplitude.

\end{document}